\documentclass[aps,twocolumn,prb,tightenlines,floatfix,showpacs]{revtex4}
\usepackage[dvips]{graphicx}
\usepackage[english]{babel}
\usepackage{amsmath}
\usepackage{amssymb}
\usepackage{tensor}

\newcommand{\vk}{\mathbf{k}}
\newcommand{\vp}{\mathbf{p}}
\newcommand{\vq}{\mathbf{q}}
\newcommand{\vx}{\mathbf{x}}
\newcommand{\vy}{\mathbf{y}}

\newcommand{\be}{\begin{eqnarray}}
\newcommand{\ee}{\end{eqnarray}}
\newcommand{\p}{\partial}

\newcommand{\psid}{\psi^{\dagger}}

\begin{document}

\title{Establishing Conservation Laws in Pair Correlated Many Body theories: T matrix Approaches}
\author{Yan He and K. Levin}
\affiliation{James Franck Institute and Department of Physics,
University of Chicago, Chicago, Illinois 60637, USA}

\begin{abstract}
We address conservation laws associated
with current, momentum and energy and show how they can be satisfied within
many body theories which focus on pair correlations.
%Our premise is that satisfaction of conservation laws in
%approximate many body theories provides an important
%validation of the approximations.
Of interest are two well known t-matrix theories which represent many
body theories which incorporate pairing in the normal state.
The first of these is associated with Nozieres Schmitt-Rink theory, while
the second involves the t-matrix of a BCS-Leggett like state as identified by
Kadanoff and Martin.
T-matrix theories begin with an ansatz for the single
particle self energy and are to be distinguished from
$\Phi$-derivable theories which introduce an ansatz for a
particular contribution to the thermodynamical potential. Conservation laws are equivalent
to Ward identities which we address in some detail here.
Although $\Phi$-derivable theories are often referred to as
``conserving theories", a consequence of this work is the demonstration
that these two t-matrix approaches similarly can be made
to obey all conservation laws.
Moreover, simplifying approximations in $\Phi$-derivable theories,
frequently lead to results which are incompatible with
conservation.
\end{abstract}
\maketitle

\section{Introduction}

In this paper we consider approximate many body fermionic theories which emphasize pairing
fluctuation effects in the normal state.
Although there are
a host of different scenarios for the centrally important ``pseudogap",
such theories are of potential interest in high temperature superconductors.
It has been
conjectured that this gap in the fermionic excitation spectrum
of the cuprates reflects pairing in advance of condensation and is to be associated with
stronger than BCS attraction \cite{Ourreview,LeggettNature}.
Even more definitively, pairing (amplitude) fluctuation theories are of interest in ultracold Fermi
gases where the interaction strength is tuneable and pairing necessarily
occurs in the normal phase \cite{Milstein}.
Among the topics of particular current interest in these cold gases
are transport phenomena.
In this context, recent attention
has focused on the
shear viscosity both experimentally \cite{Thomas} and theoretically
\cite{HaoPaper,Schaefer,Son1,NJOP}.
However, no calculation of a transport property can be considered meaningful \cite{Baym1}
without establishing conservation laws.

The goal of the present paper is to establish what are the requirements for
arriving at an approximate ``conserving theory" in the context of transport.
Approximate many body theories of pairing correlations
are of two types: either one begins with an
ansatz for the self energy [t-matrix approach]
or an ansatz for a component of the thermodynamical potential [``$\Phi$-derivability"
approach]. The latter category is more frequently associated with
``conserving theories" \cite{Baym1,Baym2}
and characterized as such in the literature. Here we emphasize that $\Phi$-derivability
is a sufficient but not necessary condition for arriving at a proper conserving theory.
Moreover,
the $\Phi$-derivability conditions are often only approximately satisfied so that
conservation laws
cannot be proved to hold.
Alternative theories known as
t-matrix theories
are the simplest category of pairing many body theories. Here one
incorporates pairing effects between the fermions by considering the summation of a series of ladder diagrams in the particle -particle channel which then
feeds back into a fermionic self energy.
In this paper we consider the simplest t-matrix approach of Nozieres Schmitt-Rink
theory \cite{NSR,OurAnnals} as well as the t-matrix introduced by Kadanoff and Martin \cite{KM} which
includes more interaction effects and
is chosen to be appropriate to BCS theory and its BCS-Leggett generalizations
\cite{Leggett,Ourreview}.
We show here
how to arrive at a proper conserving t-matrix-based transport theory.

In contrast to t-matrix schemes, the emphases of $\Phi$-derivable theories is more directly
on including multiple classes of many body diagrams, which are subject to
internal consistency. This approach is based on the observation \cite{Baym2}
that
if one starts with a contribution to the thermodynamical
potential, $\Phi$, of a certain form,
then conservation laws follow.
More specifically, consistency is represented by a key equation relating the self energy $\Sigma$
to $\Phi$.
An additional consequence of precise $\Phi$-derivability beyond the
implications for transport, is that such theories, when specialized
to the equilibrium case, obey the integrated form of conservation laws.
Nevertheless, often
there are approximations
or truncations involved so that one may violate conservation laws.
In principle, then, conservation law
tests should also be applied to \textit{approximate} $\Phi$-derivable theories
given
there is no guarantee that these approaches are fully consistent.

\subsection{Overview of Ward Identities (WI)}

In this sub-section we provide a brief overview of conservation laws or
the equivalent  Ward identities which form the basis for validating
many body theories and the basis for the present paper.
Our goal here is to introduce some of the concepts
and notation which are later addressed in more detail.
Here
we envision systems subject to an external perturbation.
The
conservation laws of interest are local conservation laws:
\be
&&\frac{\p\rho}{\p t}+\nabla\cdot{\bf J}=0,\\
&&\frac{\p{\bf J}}{\p t}+\nabla_iT_{ij}=0,\\
&&\frac{\p\rho_{\epsilon}}{\p t}+\nabla\cdot{\bf J}_{\epsilon}=0
\label{eq:3c}
\ee
Here $\rho$ and $\rho_{\epsilon}$ are particle number and energy density, ${\bf J}$, ${\bf J}_{\epsilon}$
and $T_{ij}$ are particle number, energy and momentum flow
current respectively.

In the language of 4-vectors the conservation laws for particle current
and the stress tensor take a more concise form
\begin{equation}
\p_{\mu}j^{\mu}=0
\end{equation}
and
\begin{equation}
\p_{\mu}T^{\mu\nu}=0
\end{equation}

Here the last equation can be written
as
$\p_tT^{0j}+\p_iT^{ij}=0$.
This 4-vector notation is convenient and we will use it throughout.
In this way current and charge are combined, as are space and time
and momentum and energy.

One
can alternatively address many body systems in the absence of perturbations.
Here we integrate the above
expressions in the whole space to find
global conservation laws which must be satisfied by
many body systems in the absence of perturbations, that is, in equilibrium
\be
&&\frac{d}{dt}\int d^3x\langle{\bf J}\phi\cdots\phi\rangle=0,\\
&&\frac{d}{dt}\int d^3x\langle\rho_{\epsilon}\phi\cdots\phi\rangle=0
\ee
%
%$\p_{\mu}j^{\mu}=0$ and  $\p_{\mu}T^{\mu\nu}=0$.
Here $\phi$ is an general field operator (frequently scalar).
We will reserve the notation $\psi$ to refer to specific fermionic single
particle states.

Quite generally, Ward identities, which are the central focus here, represent
the continuity equation expressed in terms of Green's functions. They
are obtained by sandwiching the operator continuity equations
written above in various time ordered products.
These Ward Identities should be satisfied in any theory, exact or
approximate.
They are, perhaps easiest to express in the context of number
conservation.
For this case it follows from
$\frac{\p\rho}{\p t}+\nabla\cdot{\bf J}=0$
that the Ward identity is given by
\be
q_\mu \Gamma^{\mu}(k+q,k)=G^{-1}(k)-G^{-1}(k+q)
\label{eq:3a}
\ee
where $G$ is the dressed Green's function and
$\Gamma$ is the dressed vertex defined by the equation
\be
&&\langle\psi(x) J^{\mu} (z) \psi^\dagger(y)\rangle\nonumber\\
&&\equiv \int G(x,x') \Gamma^{\mu}
(x',y',z) G (y',y) d^4 x' d^4y'
\ee
%\item More generally,
%inserting the various continuity equation operators into a n-particle Green's function, one can find
% that the divergence of n-particle Green's function equals to some contact terms
%.
%\be
%\p_{\mu}\langle j^{\mu}\Psi\cdots\Psi\rangle=\textrm{contact terms},\qquad
%\p_{\mu}\langle T^{\mu\nu}\phi\cdots\phi\rangle=\textrm{contact terms}
%\ee
%YAN: WHY DISCUSS n-point FUNCTIONS HERE??

More complicated is the Ward identity associated with momentum
conservation and for that reason we extensively discuss it in the
present paper.
According to Noether's theorem, the canonical stress tensor is given by
\be
T^{\mu\nu}=\sum_a\frac{\p L}{\p(\p_{\mu}\phi_a)}\p^{\nu}\phi_a-g^{\mu\nu}L
\ee
where $L$ is the Lagrangian,
$g^{\mu\nu}$ the metric
$g_{\mu\nu}=(1,-1,-1,-1)$ and index $a$ labels different species of fields.

%This satisfies the conservation law $\p_{\mu}T^{\mu\nu}=0$.
The general Ward Identity for the correlation function of scalar field and stress tensor in co-ordinate space is\cite{Coleman}
\be
&&\p_{\mu}\langle T^{\mu\nu}(x)\phi(x_1)\cdots\phi(x_n)\rangle\nonumber\\
&&=-\sum_i\delta(x-x_i)\frac{\p}{\p x_i^{\nu}}\langle\phi(x_1)\cdots\phi(x_n)\rangle
\ee
where the right hand side contributions are often referred to
as the ``contact terms". Applying the above general expression
to the 3 point correlation function case, we have
\be
&&\p_{\mu}\langle T^{\mu\nu}(x)\psid(y)\psi(z)\rangle\nonumber\\
&&=-\delta(x-y)\frac{\p}{\p y^{\nu}}\langle\psid(y)\psi(z)\rangle
-\delta(x-z)\frac{\p}{\p z^{\nu}}\langle\psid(y)\psi(z)\rangle\nonumber
\ee
Transferring to momentum and frequency space, we find
\be
&&q_\mu\Gamma^{\mu\nu}(K+Q,K)G(K+Q)G(K)\nonumber\\
&&=k^{\nu}G(K)-(k+q)^{\nu}G(K+Q)
\ee
or
\be
q_\mu\Gamma^{\mu\nu}(K+Q,K)=k^{\nu}G^{-1}(K+Q)-(k+q)^{\nu}G^{-1}(K)\nonumber
\ee
Here we define 4-momentum as $K=(\omega,\vk)$ and $Q=(q^0,\vq)$.
This last equation, to which we shall return later in the paper,  is
the general Ward Identity (WI)
for the stress tensor vertex for both non-interacting or interacting systems.
Comparing this to the analogue (Eq.~(\ref{eq:3a}))
for the number equation
(the ``$U(1)$ current vertex") we see that for the stress tensor
% WI
%discussed above $q_\mu\Gamma^{\mu}(K+Q,K)=G^{-1}(K+Q)-G^{-1}(K)$, we find that
%for stress tensor
there are extra momentum
factors multiplying each inverse Green's function. For the interacting case,
these extra momentum factors enter is subtle ways
into the Feynman diagrams and make the establishment of WI for the
interacting case more
difficult than the $U(1)$ current WI.

\section{Comparing T-matrix theories with $\Phi$-derivable theories }

For strongly correlated systems, such as ultra-cold Fermi gases in the unitary limit,
perturbation calculations are not reliable because of the lack of small parameters. To
capture the strong fluctuations, various approximation methods have been invented, one
example is t-matrix theory.
Our interest here will be on two models for the
t-matrix in which the ladder contains one or more bare
Green's functions. The third alternative involving two dressed
Green's function in the ladder (which we refer to
as ``$GG$ theory" \cite{Haussmann}) has many different versions. They
appear generally distinct from the other two schemes and are more or less based on
$\Phi$-derivable schemes \cite{Baym2}.
%The key point of conserving approximation is that the self-energy of one particle Green's
%function is a functional derivative $\Sigma=\delta\Phi/\delta G$. Here $\Phi$ is a functional
%of full dressed Green's function $G$, which also appears as the interacting part of the free
%energy in Luttinger-Ward formalism.
%In this approximation, the current, momentum and energy
%will be conserved as in the exact theory.
%In $\Phi$-derivable theories, conservation is only guaranteed when no approximations
%are made.

We label the t-matrix-based approaches by ``$G_0G$" (associated with
Kadanoff and Martin \cite{KM}) and ``$G_0G_0$" (associated with Nozieres and Schmitt-Rink \cite{NSR}).
More specifically, in
a $G_0G$ t-matrix theory\cite{Chen-review}, the self-energy is dressed by the pair propagator as
$\Sigma(K)=\sum_P t_{pg}(P)G_0(P-K)$. The pair propagator is given by the summation of infinite ladders made by bare and full Green's functions as
\be
&&t_{pg}(P)=\frac{g}{1+g\chi(P)}\\
&&\chi(P)=\sum_{K}G_0(P-K)G(K)
\ee

The behavior in Nozieres Schmitt-Rink theory is rather similar
except that all Green's functions are bare
\be
&&t_{pg}^0(P)=\frac{g}{1+g\chi^0(P)}\\
&&\chi^0(P)=\sum_{K}G_0(P-K)G_0(K)
\ee

One might ask for the justification in considering one bare and one dressed
Green's function in the ladder series, as in the first
case. This justification \cite{KM} derives from
its equivalence to BCS-like theories.
To see this, we note that BCS theory can be viewed as
incorporating virtual non-condensed pairs which
are in equilibrium with the condensate and so have a vanishing
``pair chemical potential"; that is, their excitation spectrum
is gapless.
We may interpret the t-matrix $t_{pg}(Q)$ as simply related to the
propagator for non-condensed pairs). This t-matrix satisfies the
Hugenholtz-Pines condition in the form
%Eqs (\ref{eq:4}) and (\ref{eq:6} ) introduce the self energy which must appear in
%the fully dressed Green's function $G(K)$.
\be
%t_{pg} (Q) \equiv \frac {g} { 1 + g \int G(K) G_0(-K+Q)},~~ with~
t_{pg}(Q=0) = \infty~ \rightarrow~ \mu_{pair} =0, T \leq T_c
\label{eq:3}\\
\textrm{Moreover~since}, ~~\Sigma_{sc} (K)
%= \int t_{sc}(Q) G_0(-K+Q) ~~=~~-\int \frac {\Delta_{sc}^2}
%{T} \delta(Q) G_0 (-K+Q)
= -\Delta_{sc}^2 G_0(-K)
\label{eq:4}
\ee
one can use Eq.~(\ref{eq:3}) to re-derive the BCS gap equation
\be
&&\Delta_{sc} (T) = - U \int \Delta_{sc}(T)
\frac{(1 - 2 f (E_k))} { 2 E_k}\nonumber\\
&&\textrm{with}~E_k = \sqrt{ (\epsilon_k- \mu)^2 + \Delta_{sc}^2},~ T \leq T_c .
\label{eq:5}
\ee
In $G_0G$ theory, the self-energy reflects a dressing of one
of the propagators.
The connected part of the 2 particle Green's function is also the t-matrix under
this approximation. Higher order Green's functions can be decomposed as 1 and 2 particle
Green's functions.

In evaluating the stress tensor we introduce an effective
classical field and focus
on the associated stress tensor vertex. By
including certain vertex
corrections associated with the self-energy, we demonstrate the WI is satisfied
for this
vertex. This should also imply the WI for the 1 and 2 particle Green's functions.
While the scheme is tractable we stress
that a t-matrix
approximation is clearly oversimplified as one can see that no full
dressed
internal vertex is introduced.

By contrast, in schematic form, the central equation of a $\Phi$-derivable theory is given
by a constraint on the self energy in terms of $\Phi$
defined in terms of the thermodynamical potential $\Omega $ by
\be
\Omega=\mbox{tr}\ln(-G)-\mbox{tr}(G_0^{-1}G-1)+\Phi[G]
\ee
such that
\be
\Sigma(11')=\frac{\delta\Phi[G]}{\delta  G(11')}
\ee
Here we have introduced
shorthand notation $1\equiv({\bf x}_1,\tau_1)$, etc.
Frequently, approximations need to be made. Throughout this paper when we refer
to approximate $\Phi$-derivable theories we are not referring to given class
of diagrams chosen to represent the thermodynamical potential. Rather we refer
to the adoption of further approximations
\cite{Bickers,Haussmann} made within
this scheme.  Frequently these approximate theories omit some of the
terms which should
be present in the vertex function.

Without approximations, in this approach
the one-particle Green's functions satisfy
the conservation laws and, because $\Phi$ is related to the thermodynamical
potential, the two-particle Green's functions satisfy thermodynamical
consistency.
In the most general conserving approximation, $\Phi$ can be represented as
2-particle-irreducible skeleton vacuum diagrams. In practice one has to choose a
particular sub-class of diagrams and because of this truncation, not all the WI will
necessarily be satisfied.
%It should be noted that the momentum and energy conservation is proved for the
%whole system not for local conservation.
Often approximations
violate an important symmetry such as
the crossing symmetry determined by the Pauli principle \cite{Baym1,Baym2}.

In another approach proposed by de Dominicis
and Martin \cite{Dominicis}, one introduces the full dressed 2 particle scattering vertex which is determined
via the parquet equations. In this approach, the crossing symmetry is respected but it does not
guarantee the conservation laws.
Motivated by these ideas, Bickers and
Scalapino \cite{Bickers} proposed the fluctuation exchange approximation
(FLEX), which is based on a certain choice of $\Phi[G]$. This approximation has the advantage that it satisfies
the conservation laws by construction. However,
the disadvantage of FLEX is that the vertex
which satisfies Ward identities is obtained at a different
level of approximation than that at which
the self-energy is
computed. Therefore the calculation of the self-energy
is performed with the vertices which do not satisfy
Ward identities (see e.g. the discussion in Vilk and Tremblay \cite{Tremblay}).
Similar ideas were formulated by Haussmann \cite{Haussmann}.

In summary, \textit{approximate}
conserving
theories do not guarantee the satisfaction of
conservation laws. That is, not all WI are automatically satisfied. In order to respect
crossing symmetry, one has to treat the full vertex on the same footing as the self-energy.
This seems to be a rather central challenge.

\section{Momentum current WI for free gas}

The focus of this paper is the stress tensor Ward identity. To build our
understanding we begin with the non-interacting gas.
The Lagrangian for the non-interacting system
treated as a Schrodinger field is
\be
L(x)=\frac{i}{2}\psid\p_t\psi-\frac{i}{2}\p_t\psid\psi
-\frac{1}{2m}\p_i\psid\p_i\psi+\mu\psid\psi
\ee
Here we have taken a symmetric form for the time derivative.
It follows that the equation of motion is
\be
& &\frac{\p L}{\p\psi}-\p_t\frac{\p L}{\p(\p_t\psi)}-\p_i\frac{\p L}{\p(\p_i\psi)}\nonumber\\
&=&-i\p_t\psid+\frac{1}{2m}\p_i^2\psid+\mu\psid=0
\ee
which is equivalent to the Schrodinger equation for a free Fermi gas.

The components of the canonical stress tensor involving
momentum density and momentum current are given by
\be
&&T^{0j}=-(\frac{i}{2}\psid\p_j\psi-\frac{i}{2}\p_j\psid\psi)\\
&&T^{ij}=\frac{1}{2m}(\p_i\psid\p_j\psi+\p_j\psid\p_i\psi)+\delta^{ij}L
\ee
which satisfy momentum current conservation $\p_tT^{0j}+\p_iT^{ij}=0$. One can see
that in a non-relativistic theory, the momentum density is essentially
the same as the $U(1)$ current
$J^j=T^{0j}/m$.

In momentum space, if we assume $T^{\mu j}$ carries external momentum $Q$, then the bare vertices
are
\be
&&\gamma^{0j}(K+Q,K)=k^j+\frac{q^j}{2}\\
&&\gamma^{ij}(K+Q,K)=\frac{(k+q)^ik^j+(k+q)^jk^i}{2m}\nonumber\\
&&\qquad+\delta^{ij}\Big[-\frac{(\vk+\vq)\cdot\vk}{2m}+(\omega+\frac{q^0}{2})+\mu\Big]
\ee
Taking a dot product with external momentum, we find
\be
q^i\gamma^{ij}&=&\frac{(\vk+\vq)\cdot\vq k^j+\vk\cdot\vq(k+q)^j}{2m}
-\frac{(\vk+\vq)\cdot\vk}{2m}q^j\nonumber\\
& &\qquad+\mu q^j+(\omega+\frac{q^0}{2})q^j\nonumber\\
&=&k^j(\xi_{k+q}-\xi_k)-\xi_k q^j+(\omega+\frac{q^0}{2})q^j
\ee
Then it is straightforward to verify the WI for the bare vertex as
\be
q_{\mu}\gamma^{\mu j}(K+Q,K)=k^jG^{-1}_0(K+Q)-(k+q)^jG^{-1}_0(K)\nonumber\\
\ee
which is the result we cited earlier in the paper.

\subsection{Response Functions of the stress tensor: the free gas}

The physical properties of interest are the response functions.
Once one establishes the proper form for a conserving theory of
the stress tensor, it is possible to evaluate the
general stress tensor response function given by
\be
Q^{\mu j,ab}(x-y)=-i\theta(x^0-y^0)
\langle [T^{\mu j}(x),\,T^{ab}(y)]\rangle
\ee

We next explore the consequences of momentum conservation
for the stress-stress correlations.
The divergence of the stress-stress correlation function in coordinate
space is
\be
\p_{\mu}Q^{\mu j,ab}(x-y)=-i\delta(x^0-y^0)\langle [T^{0 j}(x),\,T^{ab}(y)]\rangle
\label{eq:33}
\ee
That the right hand side is not zero arises from the so-called
``contact terms" \cite{Read}
which in turn arise from the time-ordering
in the response function definition. The commutator with $T^{0 j}(x)$ will generate spatial translation
$[T^{0 j}(\vx,t),\,\psi(\vy,t)]=i\nabla_{\vy}\psi(\vy,t)\delta^3(\vx-\vy)$.
Then in momentum space, the above equation is
\be
&&q_{\mu}Q^{\mu j,ab}(Q)=\sum_K \Big[k^jG(K)-(k+q)^j G(K+Q)\Big]\nonumber\\
&&\qquad\times\gamma^{ab}(K+Q,K)
\label{qq}
\ee

For the free Fermi gas, we can directly evaluate the correlation by diagrammatic
methods
\be
& &Q_0^{\mu\nu,\rho\lambda}(Q)\nonumber\\
&=&\sum_K \gamma^{\mu\nu}(K,K+Q)G_0(K+Q)\gamma^{\rho\lambda}(K+Q,K)G_0(K)\nonumber\\
\ee

That these equations are consistent can be
confirmed by making use of the WI for the bare vertex which
yields,
\be
&&q_{\mu}Q_0^{\mu j,ab}(Q)=\sum_K q_{\mu}\gamma^{\mu j}(K,K+Q)G_0(K+Q)\nonumber\\
&&\qquad\times\gamma^{ab}(K+Q,K)G_0(K)\nonumber\\
&=&\sum_K \Big[k^jG_0(K)-(k+q)^j G_0(K+Q)\Big]\gamma^{ab}(K+Q,K)\nonumber\\
\ee
which agrees with the general result Eq.(\ref{qq}).

\subsection{A simplified momentum current vertex}

In general, the stress tensor is not uniquely defined. Different forms
for the stress tensor will lead to different forms of WI. The canonical stress tensor
contains a time derivative which will make the frequency summation quite complex. A
more convenient form for the
stress tensor can be obtained by making use of the
equation of motion to get rid of the time derivative in $T^{ij}$. Then one
finds
\be
&&T^{0j}=-(\frac{i}{2}\psid\p_j\psi-\frac{i}{2}\p_j\psid\psi)\\
&&T^{ij}=\frac{1}{2m}(\p_i\psid\p_j\psi+\p_j\psid\p_i\psi)-\delta^{ij}\frac{\p_i^2(\psid\psi)}{4m}
\ee
The corresponding vertices are given by
\be
&&\lambda^{0j}(K+Q,K)=k^j+\frac{q^j}{2}\\
&&\lambda^{ij}(K+Q,K)=\frac{(k+q)^ik^j+(k+q)^jk^i}{2m}+\delta^{ij}\frac{q^2}{4m}\nonumber\\
\ee
We refer to this representation as introducing the $\Lambda$ vertex.

For the bare $\Lambda$ vertex, one can verify that
\be
q^i\lambda^{ij}&=&\frac{(\vk+\vq)\cdot\vq k^j+\vk\cdot\vq(k+q)^j}{2m}-\frac{q^2}{4m}q^j\nonumber\\
&=&(k^j+\frac{q^j}2)(\xi_{k+q}-\xi_k)
\ee
Thus we write the WI for this $\Lambda$ vertex as
\be
&&q_{\mu}\lambda^{\mu j}(K+Q,K)\nonumber\\
&&=(k^j+\frac{q^j}2)[G^{-1}_0(K+Q)-G^{-1}_0(K)]
\ee
which is importantly different from the WI discussed earlier.

Indeed, these two representations of the stress tensors are related by
\be
&&T_{\textrm{new}}^{ij}=T_{\textrm{old}}^{ij}-\delta^{ij}\frac12
\Big[\psid(i\p_t+\frac{\nabla^2}{2m}+\mu)\psi\nonumber\\
&&+(-i\p_t+\frac{\nabla^2}{2m}+\mu)\psid\psi\Big]
\ee
Thus we have
\be
&&\p_{\mu}\Big\langle T_{\textrm{new}}^{\mu j}(x)\psid(y)\psi(z)\Big\rangle
=\p_{\mu}\Big\langle T_{\textrm{old}}^{\mu j}(x)\psid(y)\psi(z)\Big\rangle\nonumber\\
&&+\frac12\p_{j}\Big\langle\psid(x)(i\p_t+\frac{\nabla^2}{2m}+\mu)\psi(x)\psid(y)\psi(z)\Big\rangle\nonumber\\
&&+\frac12\p_{j}\Big\langle(-i\p_t+\frac{\nabla^2}{2m}+\mu)\psid(x)\psi(x)\psid(y)\psi(z)\Big\rangle
\ee
In momentum space, we find
\be
&&q_{\mu}\lambda^{\mu j}G_0(K+Q)G_0(K)=q_{\mu}\gamma^{\mu j}G_0(K+Q)G_0(K)\nonumber\\
&&+\frac{q^j}{2}\Big[G_0(K)+G_0(K+Q)\Big]
\ee
This equation connects the first and second versions of the stress tensor
Ward identities.

In the $\Lambda$ vertex
representation, the divergence of the
stress-stress correlation function is again non zero, but introduces somewhat different
``contact terms" \cite{Read}.
\be
& &q_{\mu}Q^{\mu j,ab}(Q)\nonumber\\
&=&\langle [T^{0 j}(\vq,t),\,T^{ab}(-\vq,t)]\rangle\nonumber\\
&=& \sum_{\vp,\vk}\Big(p+\frac{q}2\Big)^j\lambda^{ab}(k+q,k)\langle[c^{\dagger}_{\vp}c_{\vp+\vq},\,
c^{\dagger}_{\vk+\vq}c_{\vk}] \rangle \nonumber\\
&=& \sum_{\vk}\Big(k+\frac{q}2\Big)^j\lambda^{ab}(k+q,k)\langle c^{\dagger}_{\vk}c_{\vk}
-c^{\dagger}_{\vk+\vq}c_{\vk+\vq} \rangle
\nonumber
\ee
In the specific case of a free Fermi gas, the divergence of stress-stress correlation can be obtained by diagrammatic methods as
\be
& &q_{\mu}Q_0^{\mu j,ab}(Q)\nonumber\\
&=&\sum_K q_{\mu}\lambda^{\mu j}(K,K+Q)G_0(K+Q)\lambda^{ab}(K+Q,K)G_0(K)\nonumber\\
&=&\sum_K (k+q/2)^j\Big[G_0(K)- G_0(K+Q)\Big]\lambda^{ab}(K+Q,K)
\nonumber
\ee

which is consistent with the WI for the $\Lambda$ vertex.

\begin{figure*}[tb]
\centerline{\includegraphics[clip,width=6in]{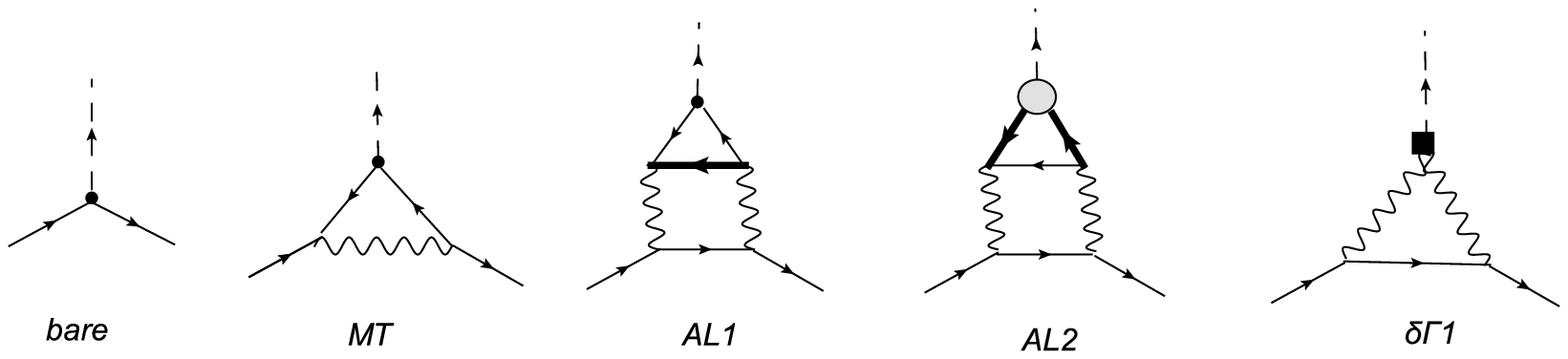}}
\centerline{\includegraphics[clip,width=6in]{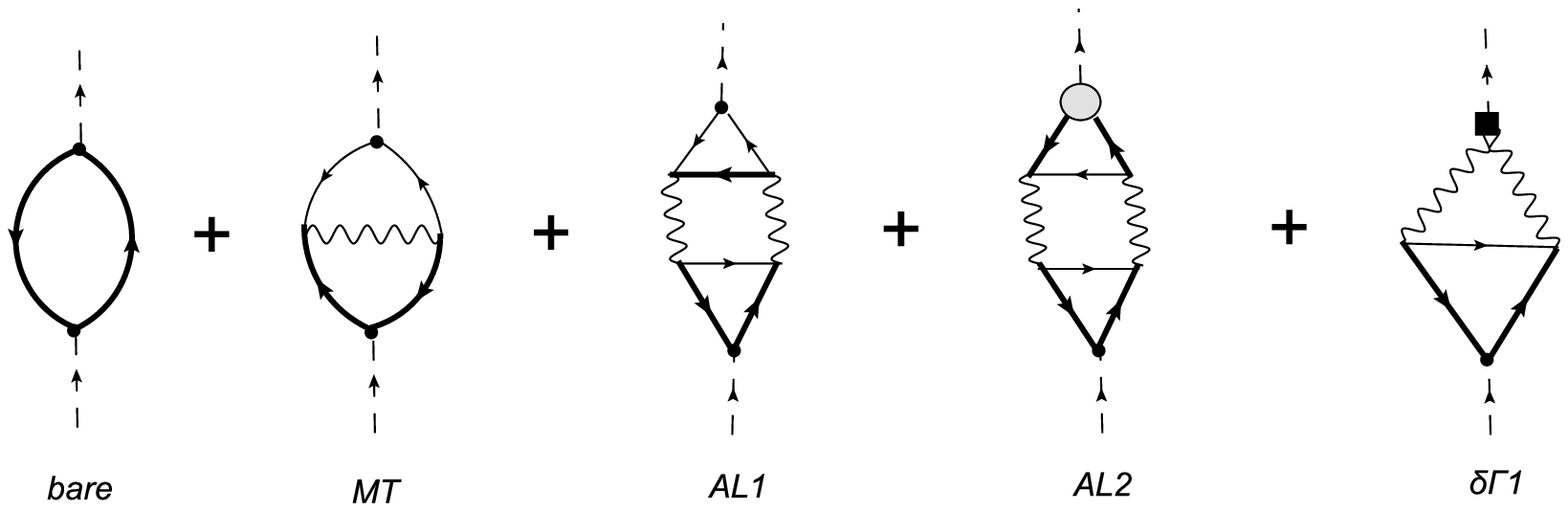}}
\caption{The diagrams contributing to the stress tensor vertex $\Gamma$.
The wiggly lines represent the $T$-matrix, thin (thick) solid lines are bare (dressed)
Green's functions, dashed lines are external stress tensor field.  The small black dots represent bare vertex
$T_0$ at zero order of $g$ and the small black square is bare vertex $T_1$ at first order of $g$. The larger
open circle represents full vertex. Labels MT, AL1 and AL2 and $\delta\Gamma_1$ are defined in text.
The lower panel is the corresponding diagrams contributing to the stress tensor correlation function.
While the results are shown for the Kadanoff Martin \cite{KM} t-matrix, one can readily
deduce the counterpart diagrams for the Nozieres Schmitt-Rink t-matrix, by assuming that
the $AL1$ and $AL2$ diagrams are equivalent.}
\label{fig:full_vertex}
\end{figure*}

\section{Stress tensor Ward Identity for interacting Fermions:
Nozieres Schmitt-Rink and $G_0G$ t-matrix theory}

We now turn to addressing the stress tensor and WI in the interacting case.
The contribution to the Lagrangian from interaction terms and the
and the equation of motion in the presence of contact interactions are
\be
&&L_{\textrm{int}}(x)=g\psid\psid\psi\psi\\
&&-i\p_t\psid+\frac{1}{2m}\p_i^2\psid+\mu\psid+2g\psid\psid\psi=0
\ee
In interacting systems, we have to introduce a new term in the stress tensor which is first order in $g$
\be
T_1^{ij}=g\delta^{ij}\,\psid\psid\psi\psi
\ee
The remaining contribution to $T^{ij}$ is the same as in
the free gas, which we refer to as $T_0$.

To verify the WI, we follow the standard
textbook approach
\cite{Peskin}. We insert the stress tensor vertex
in the self-energy diagram in all possible ways. Then for a specific
class of diagrams we can establish whether or not the WI is
satisfied.
In what follows we will present results for the more complex $GG_0$
case and note it is straightforward to extend these to the Nozieres Schmitt-Rink (NSR) case.

In order to handle the extra contribution, $T_1$
which is one order higher in $g$ than other terms, we have to insert $T_1$ into
the appropriate lower order diagrams. Since $T_1$ has four field operators, it is
sufficient to
consider only insertions into the pair propagators.

%\begin{figure}
%\centerline{\includegraphics[width=0.8\textwidth]{self-energy.eps}}
%\caption{Feynman diagram for self-energy}
%\end{figure}

\subsection{The WI for the stress tensor ($\Gamma$) vertex}

When we
insert the bare vertex $\gamma$ into the self-energy in all possible ways this
leads to three types of vertex corrections. Two of these are associated with
known literature contributions: the Aslamazov Larkin (AL) diagrams and the Maki-Thompson (MT)
diagrams which are defined as

\begin{widetext}
\be
&&\delta\Gamma^{\mu j}_{MT}(K+Q,K)=\sum_{P} t_{pg}(P)G_0(P-K-Q)\gamma^{\mu j}(P-K-Q,P-K)G_0(P-K)\\
&&\delta\Gamma^{\mu j}_{AL1}(K+Q,K)=-\sum_{P} t_{pg}(P+Q)
\Big[\sum_{P_3P_4}^{c_2}G_0(P_3+Q)\gamma^{\mu j}(P_3+Q,P_3)
G_0(P_3)G(P_4)\Big]t_{pg}(P)G_0(P-K)\nonumber\\
&&\delta\Gamma^{\mu j}_{AL2}(K+Q,K)=-\sum_{P} t_{pg}(P+Q)
\Big[\sum_{P_3P_4}^{c_2}G_0(P_3)G(P_4+Q)\Gamma^{\mu j}(P_4+Q,P_4)
G(P_4)\Big]t_{pg}(P)G_0(P-K)\nonumber
\ee
In establishing the diagram set for simple
($U(1)$) number current conservation, three types of vertex corrections are sufficient to prove
the WI. We will see that for the stress tensor WI, we need more diagrams; these come from inserting
the $T_1$ operator into the self-energy diagram.

Our derivation of the WI involves a $q$ dot product.
We write this down first for the sum of
AL diagrams which gives
\be
& &q_{\mu}\Big[\delta\Gamma^{\mu j}_{AL1}(K+Q,K)+\delta\Gamma^{\mu j}_{AL2}(K+Q,K)\Big]\nonumber\\
&=&-\sum_{P}t_{pg}(P+Q)\sum_{P_3P_4}^{c_2}
\Big[p_3^jG_0(P_3)G(P_4)+p_4^jG_0(P_3)G(P_4)\Big]t_{pg}(P)G_0(P-K)\nonumber\\
&+&\sum_{P}t_{pg}(P+Q)\sum_{P_3P_4}^{c_2}
\Big[(p_3+q)^jG_0(P_3+Q)G(P_4)+(p_4+q)^jG_0(P_3)G(P_4+Q)\Big] t_{pg}(P)G_0(P-K)\nonumber\\
&=&-\sum_{P}t_{pg}(P+Q)\Big[p^j\chi(P)-(p+q)^j\chi(P+Q)\Big]t_{pg}(P)G_0(P-K)
\ee
As noted above, $T_1$ has four legs. Thus we can directly insert it into $t_{pg}$. This
leads to
the following new correction terms as
\be
\delta\Gamma^{\mu j}_{1}(K+Q,K)=\frac{\delta^{\mu j}}{g}\sum_{P}t_{pg}(P+Q)t_{pg}(P)G_0(P-K)
\ee
There is a extra $-1/g$ factor because $T_1$ must be inserted into a lower level diagram.
Combining this with the AL diagrams, we find
\be
& &q_{\mu}\Big[\delta\Gamma^{\mu j}_{AL1}(K+Q,K)+\delta\Gamma^{\mu j}_{AL2}(K+Q,K)
+\delta\Gamma^{\mu j}_{1}(K+Q,K)\Big]\nonumber\\
&=&\sum_{P}t_{pg}(P_1+Q)
\Big[(p+q)^j t_{pg}^{-1}(P+Q)-p^j t_{pg}^{-1}(P)\Big]t_{pg}(P)G_0(P-K)\nonumber\\
&=&\sum_{P}\Big[(p+q)^j\,t_{pg}(P)G_0(P-K)-p^j\, t_{pg}(P+Q)G_0(P-K)\Big]\nonumber
\ee

For the MT diagram the dot product yields
\be
q_{\mu}\delta\Gamma^{\mu j}_{MT}(K+Q,K)
=\sum_{P}\Big[(p-k-q)^jt_{pg}(P)G_0(P-K-Q)-(p-k)^jt_{pg}(P)G_0(P-K)\Big]
\ee
Finally, collecting all terms we find the stress tensor vertex satisfies a
simple relationship (the associated Ward identity):
\be
q_{\mu}\delta\Gamma^{\mu j}(K+Q,K)
&=&\sum_{P}\Big[(k+q)^j t_{pg}(P)G_0(P-K)-k^j t_{pg}(P)G_0(P-Q-K)\Big]\nonumber\\
&=&(k+q)^j\Sigma(K)-k^j\Sigma(K+Q)
\label{eq:53}
\ee
Here $\delta\Gamma^{\mu j}=\delta\Gamma^{\mu j}_{MT}+\delta\Gamma^{\mu j}_{AL1}
+\delta\Gamma^{\mu j}_{AL2}+\delta\Gamma^{\mu j}_1$. This is the desired WI for
the two t-matrix theories under consideration.
In summary, by construction we have established a conserving diagram set for the Kadanoff-Martin \cite{KM}
$G_0G$ ladder series (as well as for the simpler NSR $G_0G_0$ t-matrix
as well). This proof depends on the fact that no further approximations are to be made
(such as a frequently used \cite{Ourreview} in a simplified self energy).
\end{widetext}

The upper panel in Figure 1 shows the right hand side of a self consistent equation for
the vertex appearing in the stress tensor correlation functions. The lower panel shows the
diagrammatic series (which includes the self consistently determined vertex) which must
be evaluated to obtain the stress tensor- stress tensor correlation. This correlation function
would enter into the shear viscosity as obtained from Eq.~(\ref{eq:68}).
While the figure is explicitly for the Kadanoff-Martin t-matrix, the results associated
with the t-matrix of Nozieres and Schmitt-Rink can be readily obtained by taking the
$AL2$ diagram to be equivalent to $AL1$. Thus, for a $G_0G_0$ t-matrix there is no
self consistency required to obtain the
corresponding vertex function.
It should be noted that, if one is interested in the shear viscosity only, a simpler
approach is to start with the current-current correlation functions as in Eq.~(\ref{eq:63}).
However, for the case of NSR theory, because there is no self consistency required,
the stress tensor correlation functions are somewhat more tractable.

\section{relation to viscosity}

Once we have a WI for the momentum flux current we have a means of evaluating
viscosity in terms of stress tensor
correlations.
One can equally well address the viscosity in terms of current-current correlations
\cite{HaoPaper}
first following Luttinger\cite{Luttinger}, and then find the correspondence
with the stress tensor correlations.

Assuming that an external vector potential
is applied to the fluid, one can readily deduce the conductivity from the Kubo formula as

\be
\sigma_{ij}(\omega)=\frac{i n}{m\omega^+}\delta_{ij}+\frac{1}{\omega^+}\int_0^{\infty}e^{i\omega t}
\lim_{\vq\to0}\langle[J_i(\vq),J_j(-\vq)]\rangle\nonumber\\
\ee
A Kubo formulation of the viscosity is, however, more subtle. Here one makes use of
the linearized
hydrodynamic equations \cite{KM2} in momentum space
\be
&&-\omega n+n_0q_i v_i=0\\
&&i\omega(\varepsilon-w_0n) \rightarrow 0 ~~~\rm{at~small~q} \\
&&-imn_0\omega v_i=n_0E_i-iq_i p\nonumber\\
&&\qquad-\Big[\eta q^2v_i+(\zeta+\frac13\eta)q_i(q_kv_k)\Big]
\label{mom}
\ee
where $n_0$, $w_0$ are the equilibrium value of density and
enthalpy per particle. We define $n$, $p$, $\varepsilon$ as fluctuations around equilibrium values of density, pressure and energy density, respectively.
These fluctuations and the velocity $v_i$ are considered as first order quantities when we linearize
hydrodynamic equations.

In the uniform or $q\to0$ limit, we have
\be
&&n=\frac{n_0q_iv_i}{\omega},\qquad v_i=-\frac{E_i}{im\omega},\qquad\varepsilon=w_0n,\nonumber\\
&&p=\Big(\frac{\p p_0}{\p n_0}\Big)_{\varepsilon}n
+\Big(\frac{\p p_0}{\p\varepsilon_0}\Big)_{n}\varepsilon=(n_0\kappa_S)^{-1}n
\ee
where $\kappa_S=n_0^{-1}(\p n_0/\p p_0)_S$ is the adiabatic compressibility.
If we substitute all the above into Eq.(\ref{mom}) and also use $J_i=n_0v_i$, we find
\be
J_i&=&\frac{in_0}{m\omega}E_i+\frac{i\kappa_S^{-1}q_i(q_kE_k)}{m^2\omega^3}\nonumber\\
& &+\frac{1}{m^2\omega^2}\Big[\eta q^2E_i+(\zeta+\frac13\eta)q_i(q_kE_k)\Big]
\ee
We can decompose any correlation function $\chi$ into transverse and longitudinal
components as
\be
\chi_{JJ}^{ij}=\chi_{JJ}^T\Big(\delta^{ij}-\frac{q^iq^j}{q^2}\Big)+\chi_{JJ}^L\frac{q^iq^j}{q^2}
\ee

If we take $q_i \perp E_i$ and $q_i\parallel E_i$ respectively, we find shear and bulk viscosity as
\be
&&\eta=\lim_{\vq\to0}\frac{m^2\omega}{q^2}i\chi_{JJ}^T, \label{eq:63}\\
&&\zeta+\frac43\eta=\lim_{\vq\to0}\frac{m^2\omega}{q^2}i\chi_{JJ}^L-\frac{i\kappa_S^{-1}}{\omega}
\ee

From the momentum flux current WI, we have $\p_t T^{0j}+\p_i T^{ij}=0$ and $T^{0i}=mJ^i$. Thus we find
an important relation between the current current and stress tensor correlation functions which
is given by
\be
m^2\omega^2\chi_{JJ}^{ij}=q^aq^b\chi_{TT}^{ia,jb}+mq^b\langle[J^i(\vq),T^{jb}(-\vq)]\rangle
\label{eq:60}
\ee
Eq.(\ref{eq:60}) is more subtle than one might have inferred owing to the
extra commutator, which is sometimes ignored in the literature.

If we introduce the viscosity tensor
\be
\eta_{ia,jb}=\eta(\delta_{ij}\delta_{ab}+\delta_{ib}\delta_{aj}
-\frac23\delta_{ia}\delta_{jb})+\zeta\delta_{ia}\delta_{jb}
\ee
then we have
\be
\eta_{ia,jb}\frac{q^aq^b}{q^2}
&=&\lim_{\vq\to0}\left(\frac{q^aq^b}{\omega q^2}\chi_{TT}^{ia,jb}
+\frac{mq^b}{\omega q^2}\langle[J^i(\vq),T^{jb}(-\vq)]\rangle\right)\nonumber\\
& &\qquad-\frac{i\kappa_S^{-1}}{\omega}\cdot\frac{q^iq^j}{q^2}
\ee

For arbitrary $q$, we thus have arrived at an expression for the shear viscosity in
terms of stress tensor correlation functions

\be
\eta_{ia,jb}&=&\lim_{\vq\to0}\left(\frac{\chi_{TT}^{ia,jb}}{\omega}
+\frac{m}{\omega}\frac{\p}{\p q_a}\langle[J^i(\vq),T^{jb}(-\vq)]\rangle\right)\nonumber\\
& &\qquad-\frac{i\kappa_S^{-1}}{\omega}\delta_{ia}\delta_{jb}
\label{eq:68}
\ee

We see that the shear viscosity is dependent not only on the stress tensor correlation
function but also on two additional terms involving the adiabatic
compressibility and the additional commutator (or contact terms). This expression
was derived earlier by N. Read and colleagues\cite{Read}.

\section{energy current WI}

Establishing the Ward identity associated with energy conservation is essential
for addressing transport coefficients such as thermopower and thermal conductivity.
Within the 4-vector notation
energy and energy current involve components of $T^{\mu \nu}$. These
are respectively given by
\be
&&T^{00}=\frac{1}{2m}\p_i\psid\p_i\psi-\mu\psid\psi\\
&&T^{j0}=-\frac{1}{2m}(\p_j\psid\p_t\psi+\p_t\psid\p_j\psi)
\ee
From Eq.~(\ref{eq:3c}), they are inter-connected through
the conservation law $\p_tT^{00}+\p_jT^{j0}=0$. In computing a thermal response,
one needs the bare and dressed vertices. The bare vertex is
given by
\be
&&\gamma^{00}(K+Q,K)=\frac{(\vk+\vq)\cdot\vk}{2m}-\mu\\
&&\gamma^{j0}(K+Q,K)=\frac{1}{2m}[(\Omega+\omega)k^j+\omega(k+q)^j]
\ee
which can be shown to satisfy
\be
q_{\mu}\gamma^{\mu 0}(K+Q,K)&=&\xi_k\Omega-\omega(\xi_{k+q}-\xi_k)\nonumber\\
&=&\omega G^{-1}_0(K+Q)-(\omega+\Omega)G^{-1}_0(K)
\nonumber
\ee

This, then presents a form of ``template" for the form of the Ward identity
associated with energy conservation
in the dressed vertex.

\subsection{Energy current WI for the interacting case: t-matrix theory}

We can prove the energy current WI in a very similar way as was done for
%the $\Gamma$ vertex of
the momentum current WI.
Here the vertex $\gamma^{\mu j}$ will be replaced by $\gamma^{\mu 0}$. For the sum of AL diagrams, we have
\be
& &q_{\mu}\Big[\delta\Gamma^{\mu 0}_{AL1}(K+Q,K)+\delta\Gamma^{\mu 0}_{AL2}(K+Q,K)\Big]\nonumber\\
&=&-\sum_P t_{pg}(P+Q)\Big[p^0\chi(P)-(p+q)^0\chi(P+Q)\Big]\nonumber\\
& &\qquad\times t_{pg}(P)G_0(P-K)
\ee
\begin{widetext}
Due to the interactions, there is an extra term $T^{\mu 0}_1=g\delta^{\mu 0}\psid\psid\psi\psi$,
in the energy current, just as there is for the momentum current.
The resulting diagrams effectively introduce insertions into $t_{pg}$.

\be
\delta\Gamma^{\mu 0}_{1}(K+Q,K)
=\frac{\delta^{\mu 0}}{g}\sum_P t_{pg}(P+Q)t_{pg}(P)G_0(P-K)
\ee
Combining this with the AL diagrams, we find
\be
& &q_{\mu}\Big[\delta\Gamma^{\mu 0}_{AL1}(K+Q,K)+\delta\Gamma^{\mu 0}_{AL2}(K+Q,K)
+\delta\Gamma^{\mu 0}_{1}(K+Q,K)\Big]\nonumber\\
&=&\sum_P \Big[(p+q)^0\,t_{pg}(P)G_0(P-K)-p^0\, t_{pg}(P+Q)G_0(P-K)\Big]
\ee

For the MT diagrams we have
\be
q_{\mu}\delta\Gamma^{\mu 0}_{MT}(K+Q,K)=\sum_P t_{pg}(P)\Big[(p-k-q)^0G_0(P-K-Q)
-(p-k)^0 G_0(P-K)\Big]
\ee
Collecting all results we find
\be
q_{\mu}\delta\Gamma^{\mu 0}(K+Q,K)&=&
\sum_P\Big[(k+q)^0\,t_{pg}(P)G_0(P-K)-k^0\, t_{pg}(P)G_0(P-K-Q)\Big]\nonumber\\
&=&(k+q)^0\Sigma(K)-k^0\Sigma(K+Q)
\ee

Here $\delta\Gamma^{\mu 0}=\delta\Gamma^{\mu 0}_{MT}+\delta\Gamma^{\mu 0}_{AL1}
+\delta\Gamma^{\mu 0}_{AL2}+\delta\Gamma^{\mu 0}_1$.
This equation which is closely analogous to
Eq.~(\ref{eq:53}) for the momentum current Ward identity is
the desired WI  for energy conservation.

\end{widetext}

\section{conclusion}

In this paper we have examined the requirements for arriving at a consistent theory
of transport. As noted in seminal earlier work \cite{Baym1}, ``In describing transport it is vital
to build the conservation laws of number, energy, momentum and angular momentum into
the structure of the approximation used to determine the thermodynamic many-particle
Green's functions."
As a sequel to this earlier study, Baym \cite{Baym2} was led to formulate
a $\Phi$-derivable theory.
What we have emphasized here is that $\Phi$-derivability is sufficient,
but not necessary. When
approximations are made to the self consistency conditions within this scheme,
there is no guarantee that a theory is conserving.
A somewhat more tractable approach, which we apply here, is to begin with an ansatz
for the self energy. Here we take as an example a
t-matrix theory which incorporates pairing
fluctuations relevant to the ultracold Fermi gases, and more general strongly
correlated superconductors and superfluids.
We have considered
two
simple t-matrix theories, that of Nozieres and Schmitt-Rink and the more self consistent
t-matrix of Kadanoff and Martin \cite{KM}.

We show how
to construct the diagrammatic series for the response functions in order to be consistent
with local conservation laws, via Ward identities.
These Ward identities become particularly complicated and not as well
known to the condensed matter community for the case of momentum conservation
(which relates to the viscosity calculations). This is the reason we have devoted more
attention to the stress tensor here.
Nevertheless,
we have
addressed local number, and energy conservation Ward identities as well.
The central finding of this work was the demonstration that these t-matrix theories,
which are not $\Phi$-derivable, are indeed
``conserving" as required for a consistent theory of transport.

\vskip8mm
We thank A. Rancon for his reading of the manuscript.
This work is supported by NSF-MRSEC Grant
0820054. We thank Hao Guo and Chih-Chun Chien for
valuable insights.

\appendix

\section{The stress tensor WI for $\Lambda$ vertex}

It is often more convenient to work with the $\Lambda$ vertex introduced in
the text.
While the Ward identities of the bare vertices are different, we will
see that the same set of diagrams are also sufficient to show the WI is satisfied for
the $\Lambda$ vertex.
Noteably
the first order vertex term $T_1$ is different from the previous case,
but the Lagrangian and equation of motion in the presence of interactions are the same.
The equation of motion for $T^{ij}$ is
\be
&&T^{ij}=\frac{1}{2m}(\p_i\psid\p_j\psi+\p_j\psid\p_i\psi)\nonumber\\
&&-\delta^{ij}\Big(\frac{\p_i^2(\psid\psi)}{4m}+g\psid\psid\psi\psi\Big)
\ee
The interaction vertex
\be
T_1^{ij}=-g\delta^{ij}\psid\psid\psi\psi
\ee
has a different sign, as compared to its counterpart.

By inserting the bare vertex $\lambda$ into the t-matrix self-energy, we still find three types of vertex corrections $\delta\Lambda^{\mu j}_{AL1}$, $\delta\Lambda^{\mu j}_{AL2}$ and $\delta\Lambda^{\mu j}_{MT}$ which are the same as before but with $\Gamma$ vertex replaced by $\Lambda$ vertex. Now the $q$ dot product with the sum of AL diagrams gives

\begin{widetext}
\be
& &q_{\mu}\Big[\delta\Lambda^{\mu j}_{AL1}(K+Q,K)+\delta\Lambda^{\mu j}_{AL2}(K+Q,K)\Big]\nonumber\\
&=&-\sum_{P}t_{pg}(P+Q)\sum_{P_3P_4}^{c_2}
\Big[(p_3+\frac q2)^jG_0(P_3)G(P_4)+(p_4+\frac q2)^jG_0(P_3)G(P_4)\Big]t_{pg}(P)G_0(P-K)\nonumber\\
&+&\sum_{P}t_{pg}(P+Q)\sum_{P_3P_4}^{c_2}
\Big[(p_3+\frac q2)^jG_0(P_3+Q)G(P_4)+(p_4+\frac q2)^jG_0(P_3)G(P_4+Q)\Big]t_{pg}(P)G_0(P-K)\nonumber\\
&=&-\sum_{P}t_{pg}(P+Q)\Big[(p+q)^j\chi(P)-p^j\chi(P+Q)\Big]t_{pg}(P)G_0(P-K)
\ee
By inserting $T_1$ into $t_{pg}$, we find
\be
\delta\Lambda^{\mu j}_{1}(K+Q,K)=-\frac{\delta^{\mu j}}{g}\sum_{P}t_{pg}(P+Q)t_{pg}(P)G_0(P-K)
\ee
which is the same as before except for an overall sign change. This term can be rewritten in two different ways as
\be
\delta\Lambda^{\mu j}_{1}(K+Q,K)&=&-\delta^{\mu j}\sum_P\Big[1-t_{pg}(P+Q)\chi(P+Q)\Big]t_{pg}(P)G_0(P-K)\nonumber\\
&=&-\delta^{\mu j}\sum_P\Big[1-t_{pg}(P)\chi(P)\Big]t_{pg}(P+Q)G_0(P-K)\nonumber
\ee
Taking the average of the above two equations, we have
\be
\delta\Lambda^{\mu j}_{1}(K+Q,K)&=&\delta^{\mu j}
\Big[t_{pg}(P+Q)\frac{\chi(P+Q)+\chi(P)}2t_{pg}(P)G_0(P-K)-\frac{\Sigma(P+Q)+\Sigma(P)}2\Big]\nonumber\\
&\equiv&\delta\Lambda^{\mu j}_{1a}(K+Q,K)+\delta\Lambda^{\mu j}_{1b}(K+Q,K)
\ee
Combining $\delta\Lambda^{\mu j}_{1a}$ with the AL diagrams, we find
\be
& &q_{\mu}\Big[\delta\Lambda^{\mu j}_{AL1}(K+Q,K)+\delta\Lambda^{\mu j}_{AL2}(K+Q,K)
+\delta\Lambda^{\mu j}_{1a}(K+Q,K)\Big]\nonumber\\
&=&\sum_{P}t_{pg}(P_1+Q)
(p+\frac q2)^j\Big[\chi(P+Q)-\chi(P)\Big]t_{pg}(P)G_0(P-K)\nonumber\\
&=&\sum_{P}(p+\frac q2)^j\Big[t_{pg}(P)G_0(P-K)-t_{pg}(P+Q)G_0(P-K)\Big]\nonumber\\
&=&\sum_{P}\Big[(p+\frac q2)^j\,t_{pg}(P)G_0(P-K)-(p-\frac q2)^j\, t_{pg}(P)G_0(P-Q-K)\Big]
\ee
For the MT diagrams we have
\be
q_{\mu}\delta\Lambda^{\mu j}_{MT}(K+Q,K)
=\sum_{P}(p-k-\frac q2)^j\Big[t_{pg}(P)G_0(P-K-Q)-t_{pg}(P)G_0(P-K)\Big]
\ee
Collecting all result we find
\be
& &q_{\mu}\Big(\delta\Lambda^{\mu j}_{AL1}+\delta\Lambda^{\mu j}_{AL2}+\delta\Lambda^{\mu j}_{1a}
+\delta\Lambda^{\mu j}_{MT}+\delta\Lambda^{\mu j}_{1b}\Big)(K+Q,K)\nonumber\\
&=&\sum_{P}\Big[(k+q)^j t_{pg}(P)G_0(P-K)-k^j t_{pg}(P)G_0(P-Q-K)\Big]
-\frac{q^j}{2}[\Sigma(P+Q)+\Sigma(P)]\nonumber\\
&=&(k+\frac q2)^j[\Sigma(K)-\Sigma(K+Q)]
\ee
Combining this equation with the bare WI, we find our desired WI for the t-matrix self-energy.
\be
q_{\mu}\Lambda^{\mu j}(K+Q,K)=(k+\frac q2)^j[G^{-1}(K+Q)-G^{-1}(K)]
\ee

\section{A simplified vertex $\lambda$ for energy current}

As in the momentum current case, we can use the
equation of motion to get rid of the time derivative
in $T^{00}$ and $T^{j0}$. In this way, we find a simplified vertex which will
make the frequency summation easier when computing the energy current response functions. The result is
\be
&&T^{00}=\frac{1}{2m}\p_i\psid\p_i\psi-\mu\psid\psi\\
&&T^{j0}=\frac{i}{2m}\Big[\p_j\psid(-\frac{\p_i^2\psi}{2m}-\mu\psi)
+(\frac{\p_i^2\psid}{2m}+\mu\psid)\p_j\psi\Big]
\ee
The bare vertex is
\be
&&\lambda^{00}(K+Q,K)=\frac{(\vk+\vq)\cdot\vk}{2m}-\mu\\
&&\lambda^{j0}(K+Q,K)=\frac{1}{2m}[(k+q)^j\xi_{k}+k^j\xi_{k+q}]
\ee
Then it can be seen that
\be
q^j\lambda^{j0}&=&\frac{1}{2m}[\vq\cdot(\vk+\vq)\xi_{k}+\vq\cdot\vk\xi_{k+q}]
=(\xi_{k+q}-\xi_k)(\xi_k+\frac{\vq\cdot\vk}{2m})
\ee
from which we obtain the WI for this bare vertex
\be
q_{\mu}\lambda^{\mu 0}(K+Q,K)=(\frac{(\vk+\vq)\cdot\vk}{2m}-\mu)
\Big[G^{-1}_0(K+Q)-G^{-1}_0(K)\Big]
\ee

We next use Eq.~(\ref{eq:33}) to write the divergence of the energy current correlation function
as
\begin{equation}
 q_{\mu}Q^{\mu 0,a0}(Q)
=\langle [T^{0 j}(\vq,t),\,T^{a0}(-\vq,t)]\rangle
\end{equation}
and evaluate the commutator as
\be
& &\langle [T^{0 0}(\vq,t),\,T^{a0}(-\vq,t)]\rangle
=\langle \sum_{\vk}\Big(\frac{(\vk+\vq)\cdot\vk}{2m}-\mu\Big)
\lambda^{ab}(k+q,k)\langle c^{\dagger}_{\vk}c_{\vk}
-c^{\dagger}_{\vk+\vq}c_{\vk+\vq} \rangle \nonumber
\ee
To validate this result, we may use diagrammatic methods to directly obtain
\be
q_{\mu}Q_0^{\mu 0,a0}(Q)&=&\sum_K q_{\mu}\lambda^{\mu 0}(K,K+Q)G_0(K+Q)\lambda^{a0}(K+Q,K)G_0(K)\nonumber\\
&=&\sum_K \Big(\frac{(\vk+\vq)\cdot\vk}{2m}-\mu\Big)\Big[G_0(K)- G_0(K+Q)\Big]
\lambda^{a0}(K+Q,K)\nonumber
\ee
which, with the WI, establishes consistency.

\end{widetext}

%\bibliographystyle{apsrev}
%\bibliography{Review2}

\end{document}